\documentclass[preprint,12pt, superscriptaddress]{revtex4}
\usepackage{graphics,epsfig}
\usepackage{epstopdf}
\usepackage{graphicx}
\usepackage{dcolumn}
\usepackage{amsmath}
\begin{document}
\title{Shadows and quasinormal modes of the Bardeen black hole in cloud of strings}

\author{Bijendra Kumar Vishvakarma}
\email{bkv1043@gmail.com}
\affiliation{Department of Physics, Institute of Science, Banaras Hindu University, Varanasi, India}
\author{Dharm Veer Singh}
\email{veerdsingh@gmail.com}
\affiliation{Department Physics,
Institute of Applied Science and Humanities, GLA University, Mathura 281406 Uttar Pradesh, 
India \footnote{Visting Associate: Inter-University Center of Astronomy and Astrophysics (IUCAA), Pune}}

\author{Sanjay Siwach}
\email{sksiwach@hotmail.com}
\affiliation{Department of Physics, Institute of Science, Banaras Hindu University, Varanasi, India}
\begin{abstract}
We investigate the  black hole (BH) solution of the Einstein's gravity coupled with non-linear electrodynamics (NED) source in the background of a cloud of strings. We analyze the horizon structure of the obtained BH solution. The optical features of the BH are explored. The photon radius and shadows  of the BH are obtained as a function of black hole parameters. We observe that the size of the shadow image is bigger than its horizon radius and photon sphere. We also study the  Quasinormal modes (QNM) using WKB formula for this black hole. The dependence of shadow radius and QN modes on black hole parameters reflects that they are mimicker to each other.
\end{abstract} 
\maketitle

\section{\label{sec:level1}Introduction}

The black holes are singular solutions of the general theory of relativity (GTR). The singularity is masked by the presence of the black hole (BH) horizon (Cosmic Censorship hypothesis). However, the singularity can be avoided in higher derivative theories of gravity itself or by coupling GTR with higher derivative matter theories \cite{8} and this can be a plausible mechanism by which nature can avoid space-time singularities \cite{1a}. Several black hole solutions of higher derivative theories exist in literature when gravity is coupled  with suitable nonlinear electrodynamics (NED) source e.g.  Born-Infeld charged black hole (BH) \cite{Jafarzade:2020ova}, Culetu BH in the presence of  cosmic string \cite{Singh:2022ycn}, regular Einstein Gauss-Bonnet (EGB) black holes \cite{Singh:2022dth}, charged $AdS$ \cite{Belhaj:2020nqy} and charged $AdS$ EGB black hole \cite{EslamPanah:2020hoj}. Bardeen black hole happens to be a regular solution and it can be obtained as a charged solution of GTR coupled with NED \cite{abg11,abg}. In fact, it is a magnetically charged solution resulting from a self-gravitating monopole. There exist many other BH solutions based on the Bardeen model
\cite{bar,fr1,Singh:2017qur,bambi,Ahmed:2022qge,sh11,Singh:2017bwj,rizzo06,
Ghosh:2020tgy,Ghosh:2020ijh,Singh:2020nwo,Singh:2019wpu,Singh:2020rnm,Singh:2021nvm,Maluf:2018lyu,ma14,dvs19,sabir,Bronnikov:2000vy,Simpson:2019mud,Panotopoulos:2019qjk,Singh:2022xgi,Singh:2020xju,bhu}.

Investigation of particle motion in black hole space-time and shadows thereof provides an interesting laboratory to test Einstein's General theory of relativity. The shadow size depends on black hole mass and parameters like charge and angular momentum. At the same time, black holes have characteristic oscillation frequencies known as Quasi-normal modes (QNM), and these modes are dependent on black hole parameters. The dependence seems to have a direct correspondence with the dependence of shadow radius and QNM \cite{Tsukamoto:2014tja,Li:2021zct,Cuadros-Melgar:2020kqn,moura2021} with BH parameters. These features are investigated recently for several types of BH's with asymptotic flat or $AdS$ space-time \cite{Sakharov:1966,Gliner:1966}. 

In this paper, we consider a BH space-time that is neither flat nor $AdS$ asymptotically. An example of this kind of space-time can be obtained by considering of BH solution in a cloud of cosmic strings \cite{Rodrigues:2022zph,Rodrigues:2022rfj}. Motivated by the recent astrophysical observations \cite{akima19,akima2019}, we are interested in the optical features of the  Bardeen BH's in the presence of cosmic string. This solution becomes the  Letelier solution \cite{xu19,l1,l2} in a certain limit. The cosmic strings and magnetic monopoles may have been produced amply in the early universe and it makes sense to think of Bardeen BH immersed in a cloud of strings. This kind of background is not asymptotically flat but provides a consistent gravitational solution.

Here, we consider the Bardeen BH solution in a cloud of strings (CoS) and study the shadows and QNM produced by it. We review the BH solution with CoS parameter using the NED in Sec. II and study its horizon structure. In sec. III  we study  the  BH shadows and QNM. We summarise and discuss our results in Sec. IV. 

\section{Exact solution of Bardeen black hole in a cloud of string}

The solution of Bardeen black hole solution in CoS was obtained by Rodrigues et. al \cite{Rodrigues:2022zph,Rodrigues:2022rfj} and we outline the procedure here. Let us begin with a theory of gravity in the presence of NED source and CoS parameters. The action is given by,
\begin{equation}
S=\int d^4x \sqrt{-g}\left[R+{\cal L}_{NED}+{\cal L}_{cs}\right],
\label{action}
\end{equation}
where $g$ is the metric determinant and $R$ denotes the  Ricci scalar. The action terms of NED and CoS are specified below. The equations of motion (EoM) can be obtained by the variation of the action (\ref{action}) with respect to $g_{\mu\nu}$ and $A_{\mu}$,
\begin{eqnarray}
&&R_{ab}-\frac{1}{2}\tilde g_{ab}R=^{NED}T_{\mu\nu}+^{CS}T_{\mu\nu}\label{egb2}\\
&& \nabla_{a}\left(\frac{\partial {{L(F)}}}{\partial F}F^{a b}\right)=0\qquad \text{and} \qquad \nabla_{a}(* F^{ab})=0,
\label{fe}
\end{eqnarray}
where $L(F )$ denotes the Lagrangian density of NED, taken as a function of $F=F_{\mu\nu}F^{\mu\nu})$. The source of the NED (Bardeen type) is \cite{Singh:2020xju,Singh:2017qur}

\begin{equation}
{{L(F)}}= \frac{3}{2sg^2}\left(\frac{\sqrt{2g^2F}}{1+\sqrt{2g^2F}}\right)^{\frac{5}{2}}
\label{nonl1}
\end{equation}

where $s=g/2M$, where $M$ and $g$ are the free parameters  associated with magnetic monopole charge  and mass.  The matter energy-momentum tensor (EMT) can also be obtained from the Eq. (\ref{nonl1}) and is given by,  

\begin{eqnarray}
&&  T_{\mu\nu}^{NED}=2\left[\frac{\partial {{L(F)}}}{\partial F}F_{\mu\sigma}F_{\nu}^{\sigma}-\tilde g_{\mu\nu}{{L(F)}}\right],
\label{emt}
\end{eqnarray}
The nonvanishing components of EMT are
\begin{eqnarray}
&&{T}^t_t = { T}^r_r = \frac{8M g^2}{(r^2+g^2)^{5/2}}\\
&&T^{\theta}_{\theta}=T^{\phi}_{\phi}=\frac{8M g^2 (r^2-4)}{(r^2+g^2)^{5/2}}
\label{emt1}
\end{eqnarray} 

Let us consider the CoS as a source governed by the Nambu-Goto action and is given by \cite{xu19}
\begin{equation}  
 S_{\text{NG}} =\int_{\Sigma}  \; m (-\gamma)^{-1/2}  d\lambda^{0} d\lambda^{1}= \int_{\Sigma}  \; m \left(-\frac{1}{2} \Sigma^{\mu \nu} \Sigma_{\mu \nu}\right)^{1/2}  d\lambda^{0} d\lambda^{1},
\end{equation}
where $\gamma$ is the determinant of the reduced metric and $m$ is a constant characterizes the mass of the each
string. $\lambda^{0}$  and  $\lambda^{1} $ are local time-like and space-like coordinates respectively \cite{24}.  The $\Sigma^{\mu \nu}$ is a bivector which is  written as
\begin{equation}
\label{eq:bivector}
\Sigma^{\mu \nu} = \epsilon^{a b} \frac{\partial x^{\mu}}{\partial \lambda^{a}} \frac{\partial x^{\nu}}{\partial \lambda^{b}},
\end{equation}
where $\epsilon^{a b}$ is the Levi-Civta tensor which takes the following non-zero values: $\epsilon^{0 1} = - \epsilon^{1 0} = 1$.
Using the definition, the EMT for CoS is given by \cite{xu19,Singh:2020nwo,Singh:2022ycn}
\begin{equation}
T^{\mu \nu}  = \frac{\rho \Sigma^{\mu \rho} \Sigma_{\rho}^{\phantom{\rho} \nu}}{\sqrt{-\gamma}}.
\end{equation}
where $\rho$ is the density. The nonvanishing components of the CoS are  
\begin{eqnarray}
&&T^t_t = T^r_r = \frac{a}{r^2},\\
&&T^{\theta}_{\theta}=T^{\phi}_{\phi}= \frac{a}{r^2},
\label{emt2}
\end{eqnarray}
where $a$ is a constant known as CoS parameter.

 In order to find the BH solution coupled with the CoS and NED source, one can consider the static spherically symmetric space-time, which is described by the following line element
\begin{equation}
ds^2 = -f(r) dt^2+ \frac{1}{f(r)} dr^2 + r^2 d\Omega^2,
\label{met1}
\end{equation}
with
\begin{equation}
f(r)=1-\frac{2m(r)}{r},
\label{met2}
\end{equation}
where $d\Omega^2=d\theta^2+\sin^2\theta d\phi^2$. Inserting   the Eqs. (11), (12), and (13) in Eq. (2),  the Einstein field equations become
 \begin{eqnarray}
\frac{d}{dr}m(r)  =\frac{2M g^2}{(r^2+g^2)^{5/2}}+\frac{a}{2}.
\label{eom1}
\end{eqnarray}
  Integrating the Eq. (\ref{eom1}) with respect to $r$. The (\ref{eom1}) becomes
\begin{equation}
m(r)=\frac{Mr^3}{(r^2+g^2)^{3/2}}+\frac{a}{2}r +C_1,
\end{equation}
where $C_1$ is constant to be identified with black hole mass, M. The line element of the black hole metric is obtained as \cite{Rodrigues:2022zph,Rodrigues:2022rfj},
\begin{equation}
ds^2=-\left(1-\frac{2M r^2}{(r^2+g^2)^{3/2}}-a\right)dt^2+\frac{dr^2}{\left(1-\frac{2M r^2}{(r^2+g^2)^{3/2}}-a\right)}+r^2d\Omega_2^2.
\label{bhs}
\end{equation}
 The solution (\ref{bhs}) is characterized by magnetic charge, $g$, and CoS parameter, $a$. The solution is exact and provides a new example in the presence of NED  and CoS.  In  the  limit, $g=0$, the resulting solution  reduces to the Letelier solution \cite{xu19} and it becomes Bardeen solution \cite{Singh:2017qur} in the absence of CoS source. It (\ref{bhs}) coincides  with the Schwarzschild  BH solution for $g=0$ and $a=0$.

Now, we study the nature of the BH horizon for the obtained BH solution (\ref{bhs}) when $(f(r)=0$:
  \begin{equation}
1-\frac{2M r^2}{(r^2+g^2)^{3/2}}-a=0 
\label{hor}
  \end{equation}
  
\begin{figure*}[ht]
\begin{tabular}{c c c c}
\includegraphics[width=.5\linewidth]{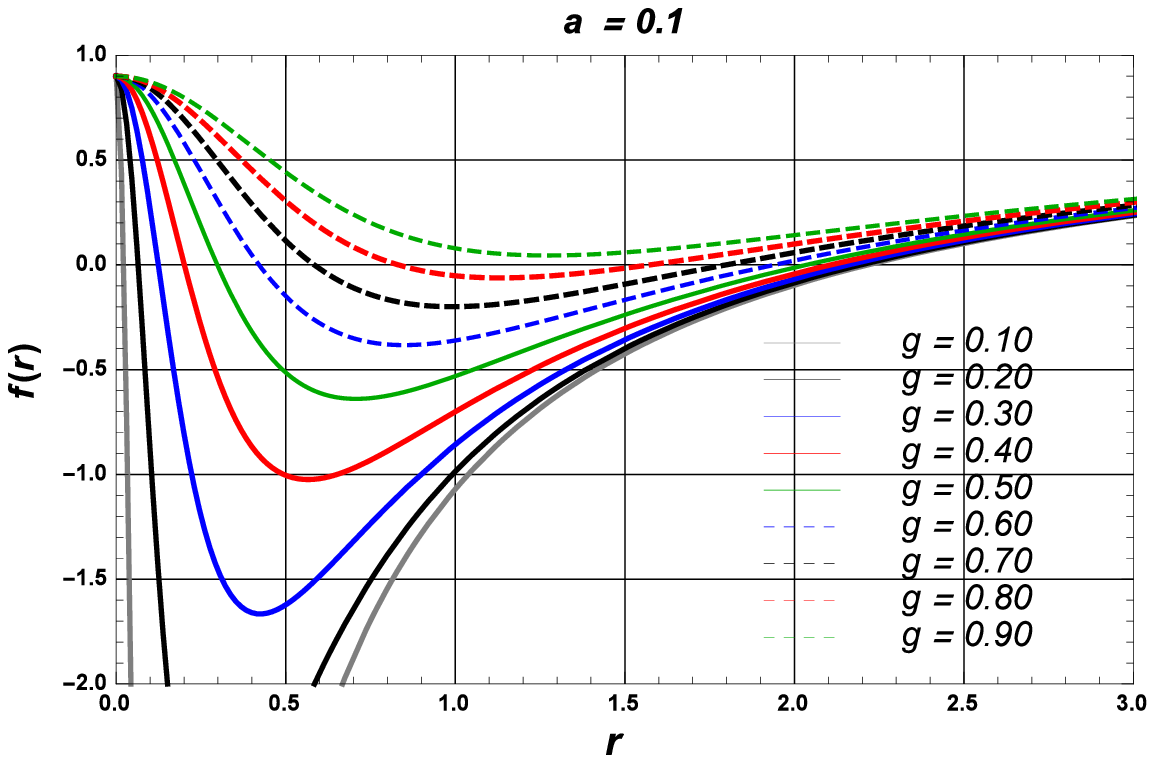}
\includegraphics[width=.5\linewidth]{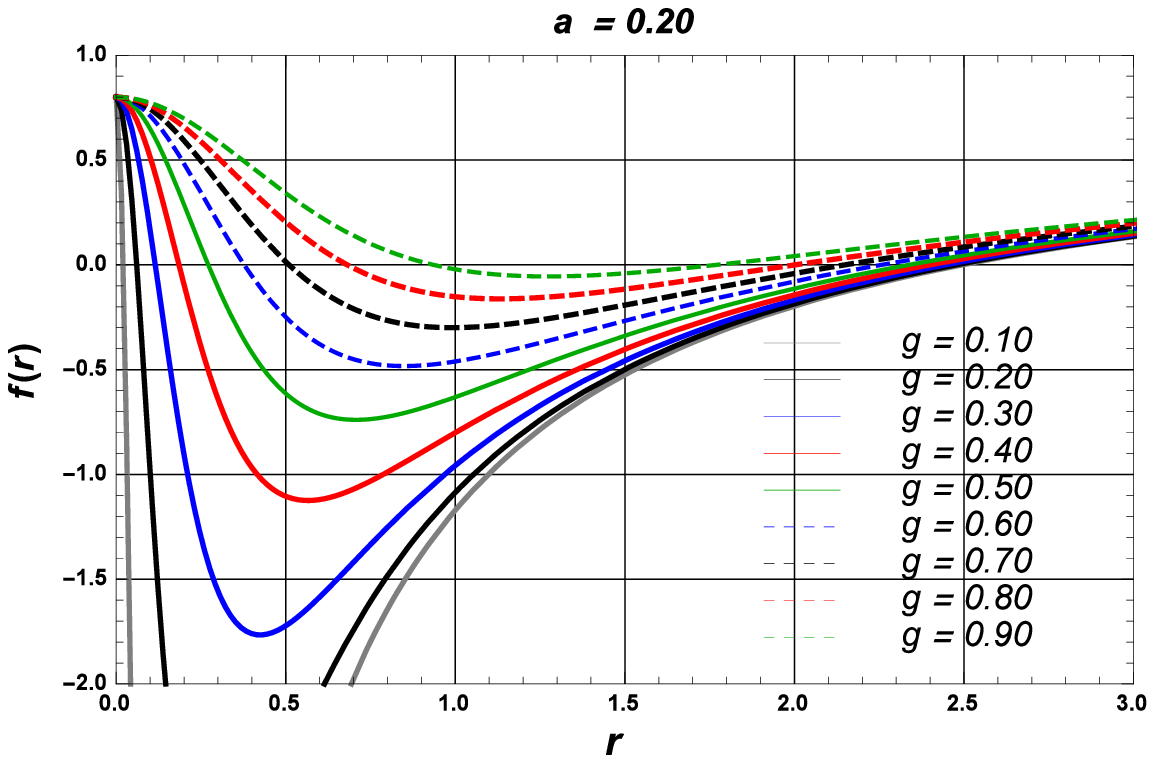}\\
\includegraphics[width=.5\linewidth]{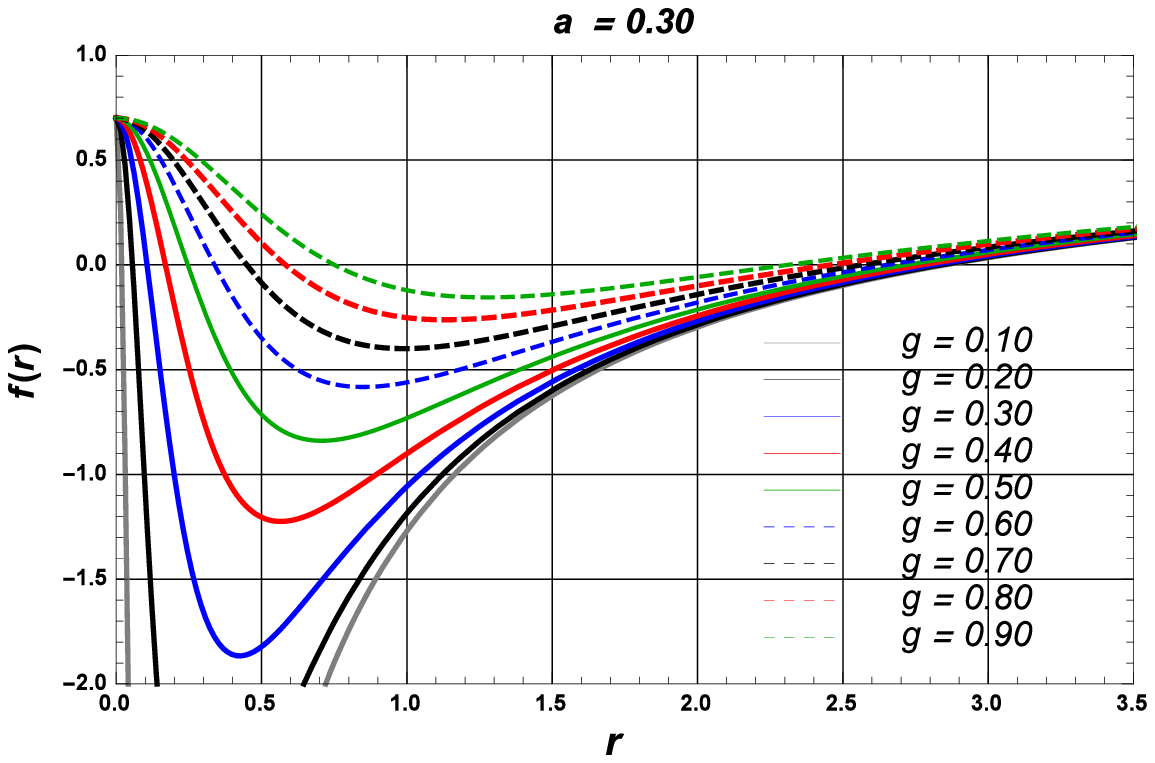}
\includegraphics[width=.5\linewidth]{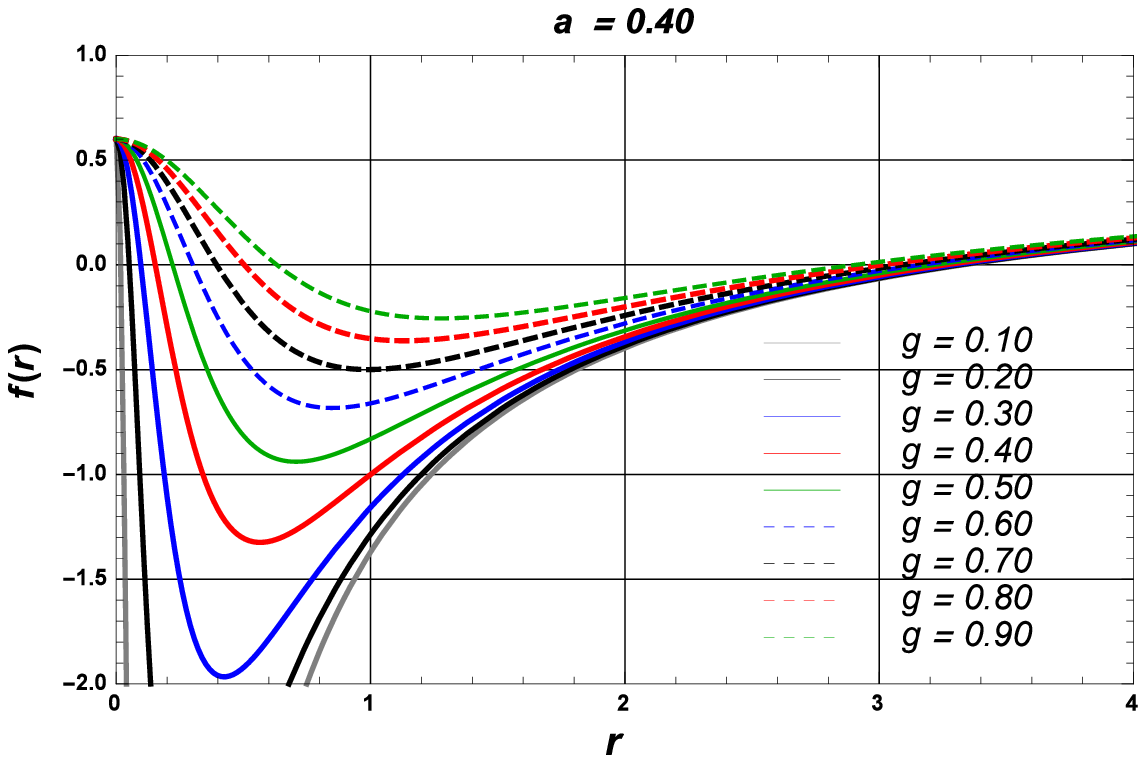}\\
\end{tabular}
\caption{Metric function $f(r)$ versus  $r$ for values of  magnetic charge $g$ and $a=0.1\,,0.2\,\, 0.3$ etc. for fixed values of $M$ and $l$.}
\label{fig:1}
\end{figure*}

The analytic solution of Eq. (\ref{hor}) does not exist and we solve it numerically in Fig.(\ref{fig:1}). The numerical values of the horizon radii are shown in table (\ref{tr3}) for different values of the CoS parameter, $a$, and magnetic charge, $g$. In Tab.  (\ref{tr3}), we see that the size of the BH horizon decreases with growing  magnetic charge, $g$, and CoS parameter, $a$. The BH solution has degenerated  at the value of $(g_c=0.85)$ and $(a=0.1)$ the BH horizon merges are  known as  the degenerate horizon. When the value of $g>0.85$, no black hole solutions  exists. The critical value $g_c$ also increases with the increase in the CoS parameter, $a$.

\begin{table}[ht]
 \begin{center}
 \begin{tabular}{ |l | l   | l   | l   |  l |  l | l | l | l | l | }
\hline
            \hline
  \multicolumn{1}{|c|}{ } &\multicolumn{1}{c}{}  &\multicolumn{1}{c}{$a=0.1$}  &\multicolumn{1}{c|}{} &\multicolumn{1}{c}{}&\multicolumn{1}{c}{$a=0.2$}&\multicolumn{1}{c|}{}&\multicolumn{1}{c}{}&\multicolumn{1}{c}{$a=0.3$}&\multicolumn{1}{c|}{}\\
            \hline
                  
  \multicolumn{1}{|c|}{ $g$} &\multicolumn{1}{c|}{$r_-$}  &\multicolumn{1}{c|}{$r_+$}  &\multicolumn{1}{c|}{$\delta$} &\multicolumn{1}{c|}{$r_-$}&\multicolumn{1}{c|}{$r_+$}&\multicolumn{1}{c|}{$\delta$}&\multicolumn{1}{c|}{$r_-$}&\multicolumn{1}{c|}{$r_+$}&\multicolumn{1}{c|}{$\delta$}\\
  \hline
            \,\,\,\,\,0.1 ~~  &~~0.020~~  & ~~2.24~~ & ~~2.22~~ &  ~~0.0207~~& ~~2.51~~& ~~2.489~~& ~~0.0276~~& ~~2.87~~&~~2.842~~ \\            
             \,\,\,\,\,0.2~~  &~~0.0622~~  & ~~2.19~~ & ~~2.12~~ & ~~0.0737~~& ~~2.47~~& ~~2.396~~& ~~0.0753~~& ~~2.84~~& ~~2.764~~ \\ 
              \,\,\,\,\,0.3 ~~  &~~0.131~~  & ~~2.17~~ & ~~2.04~~ & ~~0.110~~& ~~2.44~~& ~~2.330~~& ~~0.110~~& ~~2.82~~& ~~2.710~~ \\ 
               \,\,\,\,\,0.4 ~~  &~~0.207~~  & ~~2.11~~ & ~~1.903~~ & ~~0.193~~& ~~2.41~~& ~~2.217~~& ~~0.165~~& ~~2.76~~& ~~2.595~~ \\ 
                    \,\,\,\,\,0.5~~  &~~0.297~~  & ~~2.03~~ & ~~1.733~~ & ~~0.285~~& ~~2.35~~& ~~2.065~~& ~~0.237~~& ~~2.73~~& ~~2.493~~ \\ 
                        \,\,\,\,\,0.6 ~~  &~~0.421~~  & ~~1.94~~ & ~~1.519~~ & ~~0.359~~& ~~2.26~~& ~~1.901~~& ~~0.331~~& ~~2.66~~& ~~2.329~~ \\ 
                            \,\,\,\,\,0.7 ~~  &~~0.581~~  & ~~1.80~~ & ~~1.219~~ & ~~0.506~~& ~~2.12~~& ~~1.614~~& ~~0.451~~& ~~2.57~~& ~~2.119~~ \\ 
                                \,\,\,\,\,0.8 ~~  &~~0.829~~  & ~~1.61~ & ~~0.781~~ & ~~0.672~~& ~~2.02~~& ~~1.348~~& ~~0.571~~& ~~2.46~~& ~~1.889~~ \\ 
                 \,\,\,\,\,$g_c$ ~~  &~~1.197~~  & ~~1.197~~ & ~~0~~ & ~~1.327~~& ~~1.327~~& ~~0~~& ~~1.521~~& ~~1.521~~& ~~0~~ \\ 

            \hline 
\hline
        \end{tabular}
        \caption{The analysis of inner and outer horizon for different values of  CoS parameter, $a$ and magnetic charge, $g$ for a fixed value of $M=1$. The critical value of magnetic charge is  $g_c=0.85\,,0.97$ and $ 1.12$ corresponding to the $a=0.1\,, 0.2$ and $0.3$ respectively.}
\label{tr3}
    \end{center}
\end{table}

 It was shown by Rodriques et. al. \cite{Rodrigues:2022zph,Rodrigues:2022rfj} that the curvature invariants  like Kretschmann scalar $R_{\mu\nu\lambda\sigma}R^{\mu\nu\lambda\sigma}$ remain singular at the center for non-zero value of CoS parameter, a. This is unlike the Bardeen solution where the singularity at the center was resolved by magnetic charge.
  


\section{Shadow and Quasi normal Modes}

Let us consider the massless photon moving in the  background of the BH solution (\ref{bhs}). We consider the motion of the photon restricted to the equatorial plane by demanding $\theta=\pi/2$. The corresponding EoM can be obtained using the Hamiltonian \cite{Belhaj:2020okh,Belhaj:2022kek,Belhaj:2021tfc},
\begin{equation}
H=\frac{1}{2}\left[-\frac{p_t^2}{f(r)}+f(r)p_r^2+\frac{p_{\phi}^2}{r^2}\right].
\end{equation}
The canonically conjugate momentum for the BH metric (\ref{bhs}) can be obtained as
\begin{eqnarray}
&&p_t=\left(1-\frac{2M r^2}{(r^2+g^2){3/2}}-a\right){\dot t}=E,\qquad\qquad p_r=\left[\left(1-\frac{2M r^2}{(r^2+g^2){3/2}}-a\right)\right]^{-1}{\dot r},\nonumber\\
&&p_{\theta}=r^2{\dot\theta},\qquad\qquad\qquad\quad\quad\qquad\qquad\qquad\qquad p_{\phi}=r^2\sin^2\theta {\dot \phi}=L.
\end{eqnarray}

where $E$ and $L$ are the energy and angular momentum respectively. The EoM is obtained as
\begin{eqnarray}
&&{\dot t}=\frac{E}{\left(1-\frac{2M r^2}{(r^2+g^2){3/2}}-a\right)},\qquad\qquad r^2{\dot r}=\pm\sqrt{\cal R}\nonumber\\
&&r^2{\dot \theta}=\pm\sqrt{{ \Theta}},\qquad \qquad \qquad\qquad\quad{\dot\phi}=\frac{L}{r^2}.
\end{eqnarray}
We can re-write the radial null geodesics equation as follows
\begin{equation}
    {\dot r}^2+V_{eff}(r)=0 \qquad \text{with} \qquad V_{eff}=f(r)\left(\frac{L^2}{r^2}+\frac{E^2}{f(r)}\right) 
\end{equation}
The  null circular geodesics  satisfies the following conditions
\begin{equation}
V_{eff}=0, \qquad \text{and}\qquad \frac{\partial V_{eff}}{\partial r}=0.
\label{pot}
\end{equation}
and it gives
\begin{equation}
\frac{(1-a)(g^2+r^2_p)^{5/2}-3M r^4_p}{(g^2+r^2_p)[(1-a)(g^2+r^2_p)^{5/2}-2M r^2_p]}=0.
\label{rp}
\end{equation}
The analytic solution can not be obtained for the photon radius from this equation.  The numerical values radius of photon for different values of magnetic charge, $g$, and CoS parameter, $a$ and are listed below:

\begin{table}[ht]
 \begin{center}
 \begin{tabular}{ |l | l   | l   | l   |  l |  l | l | l | l | l | }
\hline
            \hline
  \multicolumn{1}{|c}{ } &\multicolumn{1}{c}{}  &\multicolumn{1}{c}{}  &\multicolumn{1}{c }{ }&\multicolumn{1}{c }{ }&\multicolumn{1}{c }{ $r_p$}&\multicolumn{1}{c }{ }&\multicolumn{1}{c }{ }&\multicolumn{1}{c }{ } &\multicolumn{1}{c|}{}\\
            \hline
  \multicolumn{1}{|c|}{ $a$} &\multicolumn{1}{c|}{$g=0.1$}  &\multicolumn{1}{c|}{$g=0.2$}  &\multicolumn{1}{c|}{$g=0.3$} &\multicolumn{1}{c|}{$g=0.4$}&\multicolumn{1}{c|}{$g=0.5$}&\multicolumn{1}{c|}{$g=0.6$}&\multicolumn{1}{c|}{$g=0.7$}&\multicolumn{1}{c|}{$g=0.8$}&\multicolumn{1}{c|}{$g=0.9$}\\
            \hline
            \,\,\,\,\,0.1 ~~  &~~3.325~~  & ~~3.302~~ & ~~3.263~~ & ~~3.207~~& ~~3.129~~& ~~3.027~~& ~~2.890~~& ~~2.701~~& ~~2.397~~ \\            
             \,\,\,\,\,0.2~~  &~~3.743~~  & ~~3.723~~ & ~~3.688~~ & ~~3.639~~& ~~3.572~~& ~~3.486~~& ~~2.375~~& ~~2.231~~& ~~2.039~~ \\ 
              \,\,\,\,\,0.3 ~~  &~~4.279~~  & ~~4.262~~ & ~~4.232~~ & ~~4.189~~& ~~4.132~~& ~~4.060~~& ~~3.969~~& ~~3.837~~& ~~3.039~~ \\ 
               \,\,\,\,\,0.4 ~~  &~~4.994~~  & ~~4.979~~ & ~~4.954~~ & ~~4.918~~& ~~4.870~~& ~~4.810~~& ~~4.737~~& ~~4.648~~& ~~3.716~~ \\ 
                    \,\,\,\,\,0.5~~  &~~5.995~~  & ~~5.973~~ & ~~5.962~~ & ~~5.932~~& ~~5.893~~& ~~5.844~~& ~~5.785~~& ~~5.715~~& ~~5.633~~ \\ 
                        \,\,\,\,\,0.6 ~~  &~~7.496~~  & ~~7.486~~ & ~~7.469~~ & ~~7.446~~& ~~7.415~~& ~~7.377~~& ~~7.331~~& ~~7.278~~& ~~7.216~~ \\ 
                            \,\,\,\,\,0.7 ~~  &~~9.997~~  & ~~9.989~~ & ~~9.977~~ & ~~9.959~~& ~~9.936~~& ~~9.908~~& ~~9.875~~& ~~9.836~~& ~~9.791~~ \\ 
                                \,\,\,\,\,0.8 ~~  &~~14.998~~  & ~~14.989~~ & ~~14.985~~ & ~~14.973~~& ~~14.958~~& ~~14.936~~& ~~14.917~~& ~~14.892~~& ~~14.863~~ \\ 
                 \,\,\,\,\,0.9 ~~  &~~29.999~~  & ~~29.996~~ & ~~29.992~~ & ~~29.983~~& ~~29.979~~& ~~29.970~~& ~~29.959~~& ~~29.946~~& ~~29.932~~ \\ 
                  \,\,\,\,\,0.99 ~~  &~~300~~  & ~~300~~ & ~~299.999~~ & ~~299.99~~& ~~299.999~~& ~~299.997~~& ~~299.996~~& ~~299.995~~& ~~299.999~~ \\ 
            \hline 
\hline
        \end{tabular}
        \caption{The magnitude of photon radius with the variation of  CoS parameter, $a$ and magnetic charge, $g$ for a fixed value of  $M=1$.}
\label{tr1}
    \end{center}
\end{table}
It is noticed that (Tab. \ref{tr1}) the effect of CoS parameter $a$ and monopole  charge, $g$ are opposite to each other on photon radii. the radius of the photon increases with the increase the CoS parameter, $a$ but decreases with the increase of the magnetic charge $g$.  

\subsection{Black Hole Shadow}

Let us study the behavior of the shadow radii  of the BH solution (\ref{bhs}). The size of the BH shadow can be written as \cite{72}
\begin{equation}
r_s=\frac{r_p}{\sqrt{f(r_p)}}.
\end {equation}

The numerical value of the shadow radius is given below in Tab. III and plotted in Fig. 2 for different values of BH parameters. We notice that (Tab. \ref{tr2}) the magnitude of the shadow radius increases with the increase of the CoS parameter, $a$, and decreases with the increase in magnetic charge $g$.
\begin{table}[ht]
 \begin{center}
 \begin{tabular}{ |l | l   | l   | l   |  l |  l | l | l | l | l | }
\hline
            \hline
  \multicolumn{1}{|c}{ } &\multicolumn{1}{c}{}  &\multicolumn{1}{c}{}  &\multicolumn{1}{c }{ }&\multicolumn{1}{c }{ }&\multicolumn{1}{c }{ $r_s$}&\multicolumn{1}{c }{ }&\multicolumn{1}{c }{ }&\multicolumn{1}{c }{ } &\multicolumn{1}{c|}{}\\
            \hline
  \multicolumn{1}{|c|}{ $a$} &\multicolumn{1}{c|}{$g=0.1$}  &\multicolumn{1}{c|}{$g=0.2$}  &\multicolumn{1}{c|}{$g=0.3$} &\multicolumn{1}{c|}{$g=0.4$}&\multicolumn{1}{c|}{$g=0.5$}&\multicolumn{1}{c|}{$g=0.6$}&\multicolumn{1}{c|}{$g=0.7$}&\multicolumn{1}{c|}{$g=0.8$}&\multicolumn{1}{c|}{$g=0.9$}\\
            \hline
            \,\,\,\,\,0.1 ~~  &~~6.077~~  & ~~6.052~~ & ~~6.010~~ &  ~~5.948~~& ~~5.865~~& ~~5.757~~& ~~5.617~~& ~~5.431~~&~~5.164~~ \\            
             \,\,\,\,\,0.2~~  &~~7.254~~  & ~~7.236~~ & ~~7.190~~ & ~~7.1333~~& ~~7.060~~& ~~6.960~~& ~~6.836~~& ~~6.681~~& ~~6.481~~ \\ 
              \,\,\,\,\,0.3 ~~  &~~8.865~~  & ~~8.843~~ & ~~8.806~~ & ~~8.753~~& ~~8.685~~& ~~8.595~~& ~~8.466~~& ~~8.353~~& ~~8.099~~ \\ 
               \,\,\,\,\,0.4 ~~  &~~11.173~~  & ~~11.153~~ & ~~11.119~~ & ~~11.070~~& ~~11.009~~& ~~10.928~~& ~~10.831~~& ~~10.715~~& ~~10.419~~ \\ 
                    \,\,\,\,\,0.5~~  &~~14.672~~  & ~~14.672~~ & ~~14.641~~ & ~~14.597~~& ~~14.541~~& ~~14.470~~& ~~14.384~~& ~~14.283~~& ~~14.165~~ \\ 
                        \,\,\,\,\,0.6 ~~  &~~20.534~~  & ~~20.517~~ & ~~20.490~~ & ~~20.451~~& ~~20.401~~& ~~20.338~~& ~~20.264~~& ~~20.117~~& ~~20.076~~ \\ 
                            \,\,\,\,\,0.7 ~~  &~~31.618~~  & ~~31.608~~ & ~~31.580~~ & ~~31.546~~& ~~31.503~~& ~~31.450~~& ~~31.387~~& ~~31.317~~& ~~31.229~~ \\ 
                                \,\,\,\,\,0.8 ~~  &~~58.090~~  & ~~58.079~ & ~~58.059~~ & ~~58.320~~& ~~57.997~~& ~~57.944~~& ~~57.903~~& ~~57.849~~& ~~57.777~~ \\ 
                 \,\,\,\,\,0.9 ~~  &~~164.31~~  & ~~164.30~~ & ~~164.39~~ & ~~164.27~~& ~~164.24~~& ~~164.28~~& ~~164.18~~& ~~164.14~~& ~~164.09~~ \\ 
                  \,\,\,\,\,0.99 ~~  &~~5196.1~~  & ~~5196.1~~ & ~~5196.1~~ & ~~5196.1~~& ~~5196.1~~& ~~5196.1~~& ~~5196.1~~& ~~5196.1~~& ~~5196.0~~ \\ 
            \hline 
\hline
        \end{tabular}
        \caption{The magnitude of photon radius  with the variation of  CoS parameter ($a$) and magnetic charge ($g$) for a fixed value of  $M=1$.}
\label{tr2}
    \end{center}
\end{table}
\begin{figure*}[ht]
\begin{tabular}{c c c c}
\includegraphics[width=.5\linewidth]{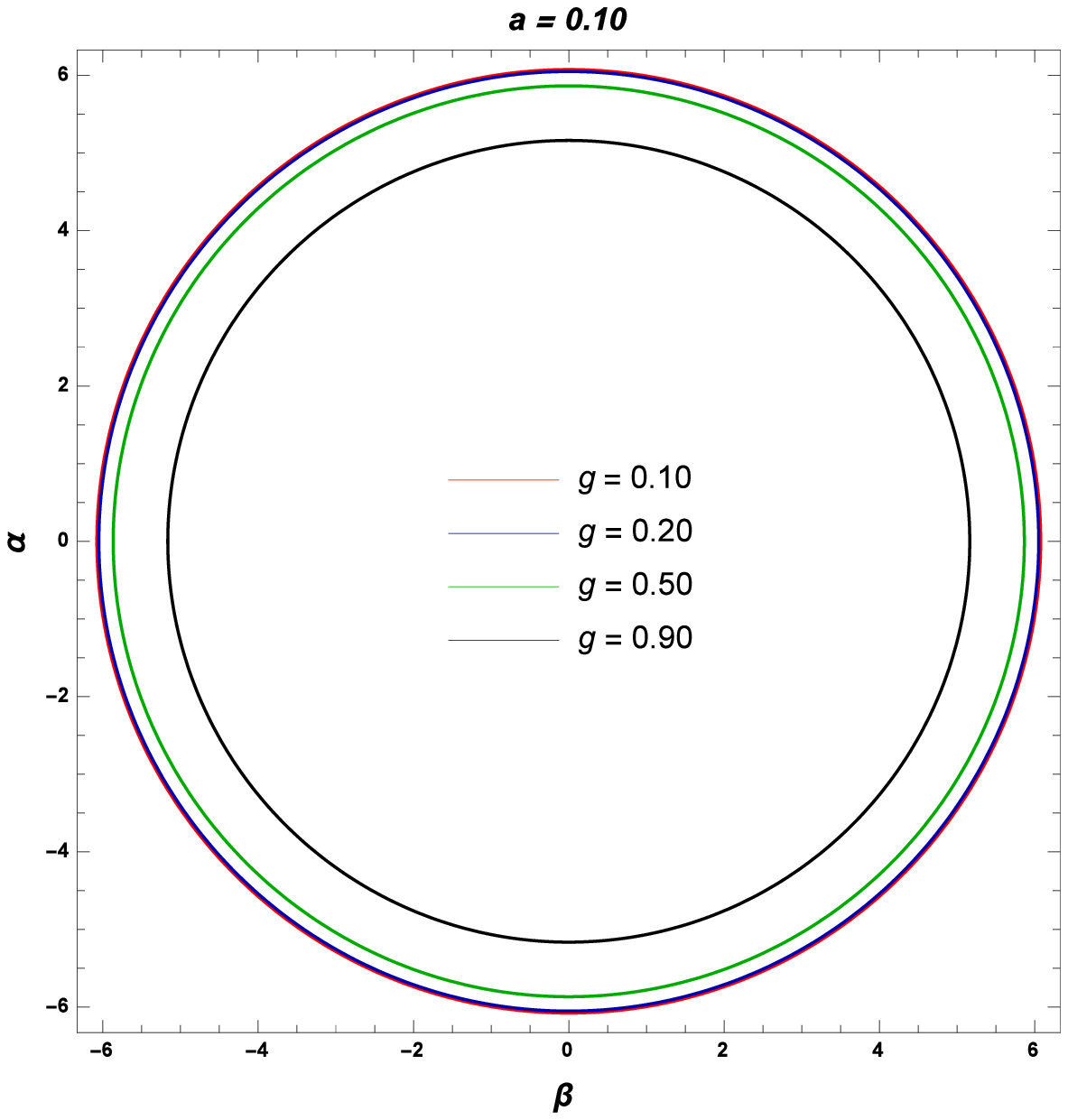}
\includegraphics[width=.5\linewidth]{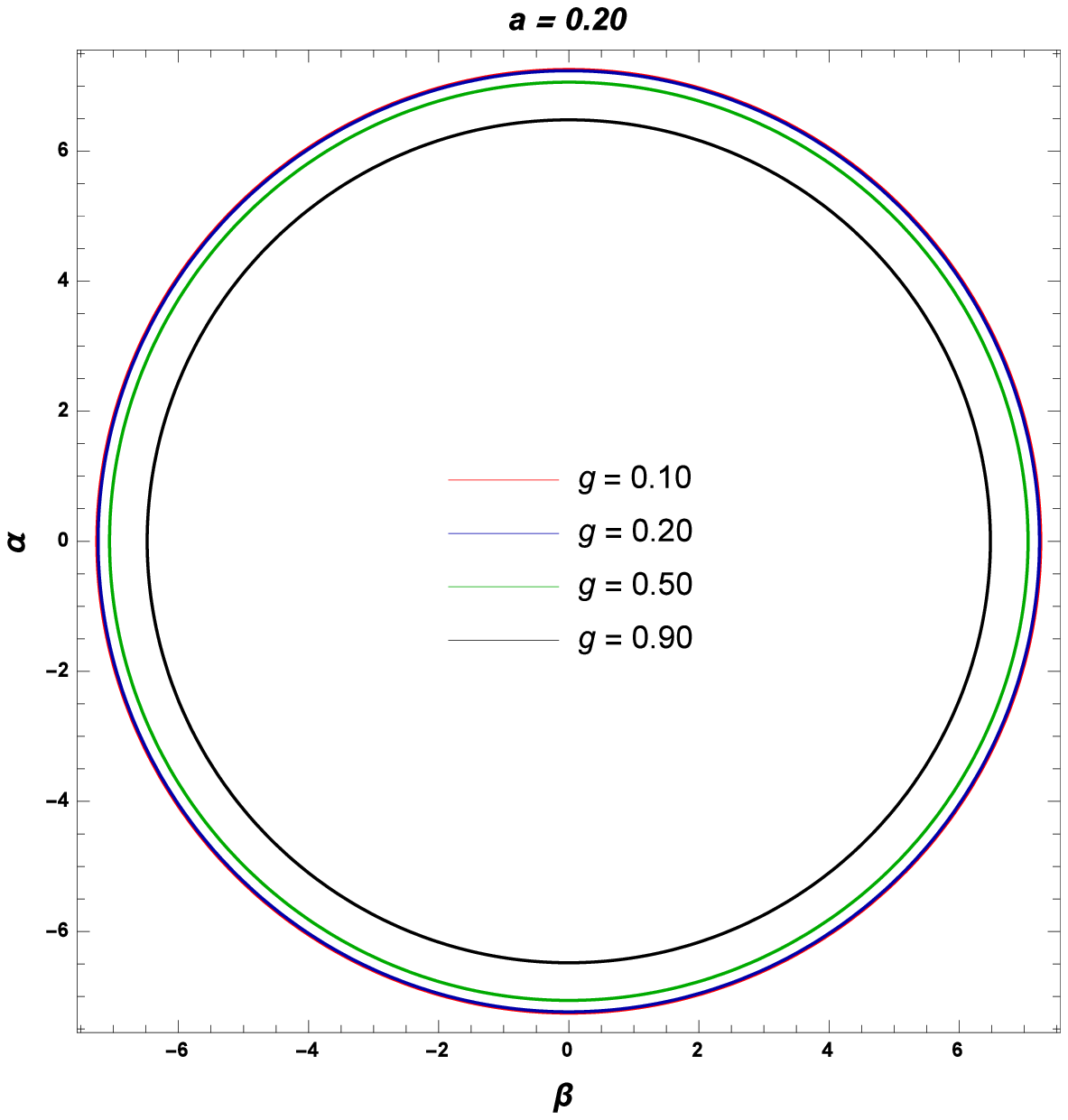}\\
\includegraphics[width=.5\linewidth]{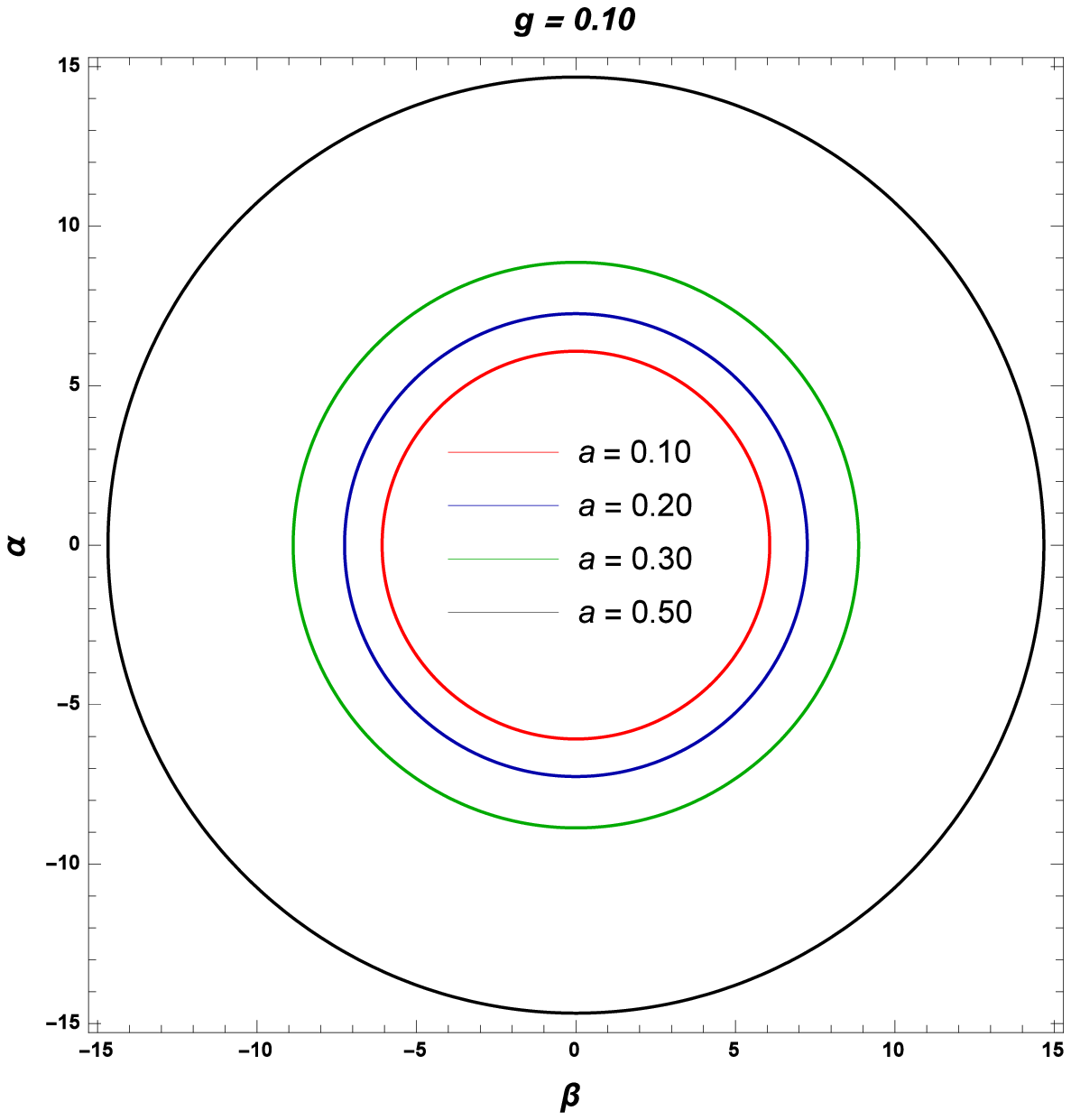}
\includegraphics[width=.5\linewidth]{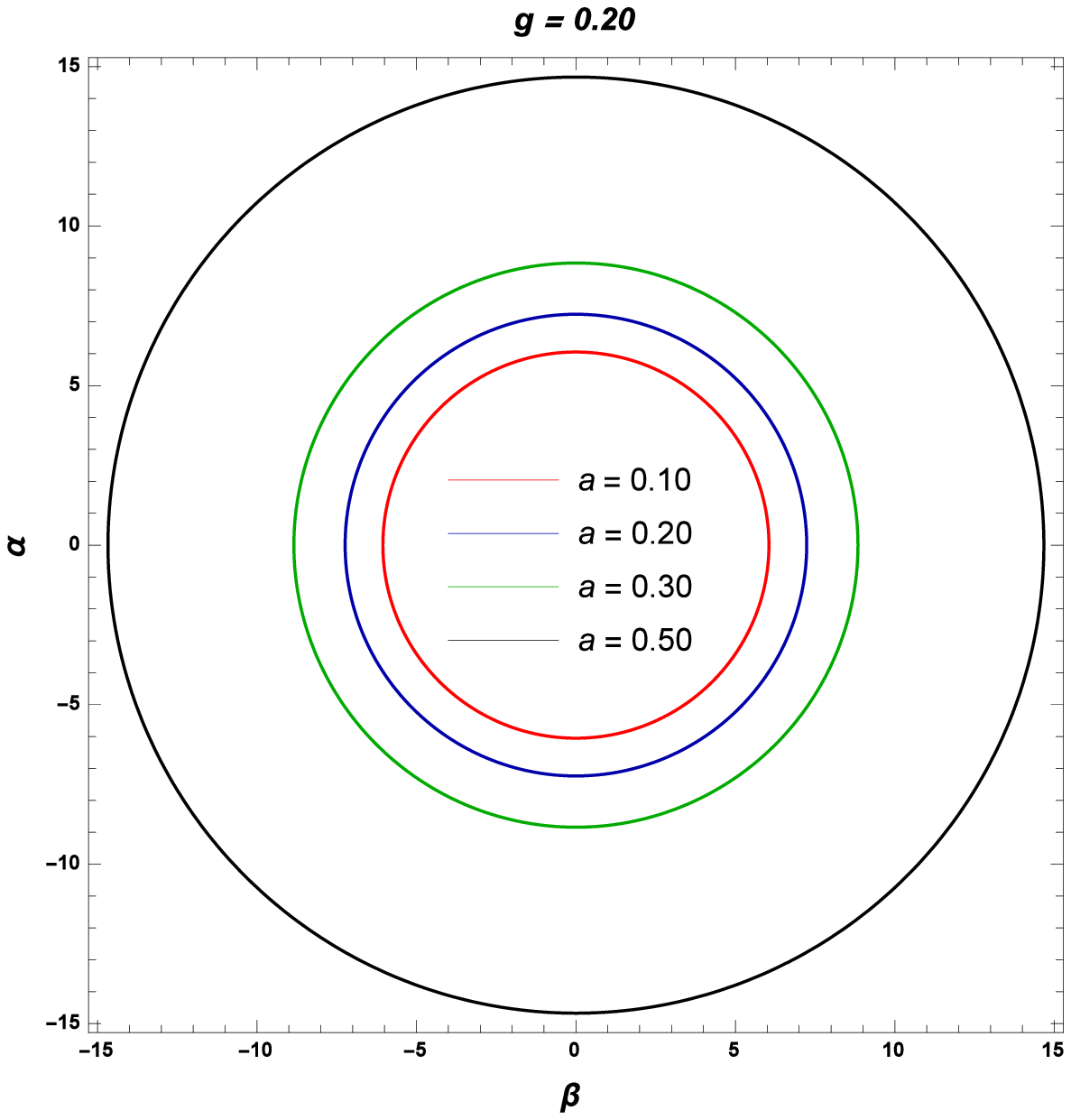}
\end{tabular}
\caption{ The  shadow of a BH for different value of global charge ($a$) and MM charge $g$ for $0.1\, 0.2\,\, 0.3$ and $a=0.4$ with fixed  $M$.}
\label{fig2}
\end{figure*}

\subsection{Quasinormal Modes}
Let us consider the QNM of the solution in order to study the dynamical stability of the obtained BH solution (\ref{bhs}). It is characterized by the real and imaginary parts of the QNM frequencies,  ($\omega=\omega_R+i \omega_I$). If $\omega>0$, the BH is unstable and $\omega<0$, it is stable. We can compute the QNMs and QNFs by solving the scalar field equation in the black hole space-time \cite{Singh:2022xgi},
\begin{equation}
\frac{1}{\sqrt{-g}}\partial_{\mu}\left(\sqrt{-g} g^{\mu\nu}\partial_{\nu}\right)\phi=0.
\label{scalar1}
\end{equation}
The solution can be obtained by separation of variables as, 
\begin{equation}
\phi=\frac{1}{r}\sum_{lm}e^{i\omega t} u_{lm}(r)Y^m_{l}(\theta,\phi),
\label{scalar2}
\end{equation}
 where $Y^m_{l}$  are  spherical harmonics. The radial equation takes the Schrodinger-like form if we use the tortoise co-ordinate $dr^{*}=dr/f(r)$
\begin{equation}
\left(\frac{d^2}{dr^{*^2}}+\omega^2-V_0(r^{*})\right) u(r)=0,
\end{equation}
where, $V_0(r^{*})=f\left(\frac{f'}{f}+\frac{l(l+1)}{r^2}\right)$.
 
The QNFs are complex numbers given by $\omega=\omega_R+i\omega_I$.  We use the WKB formula in large $l$ limit \cite{will,iyer,konoplya1,wkb1,yar} to obtain QNFs.

\begin{equation}
\omega=l\Omega-i\left(n+\frac{1}{2}\right)|\Lambda|,    
\end{equation}
 with    
\begin{eqnarray}
 \Omega=\frac{\sqrt{f(r_p)}}{r_p}=\frac{1}{L_p}\qquad \text{and}\qquad \Lambda=\frac{\sqrt{2f(r_p)-r^2_pf''(r_p)}}{\sqrt{2} L_p}.
\end{eqnarray}

\begin{figure*}[ht]
\begin{tabular}{c c c c}
\includegraphics[width=.5\linewidth]{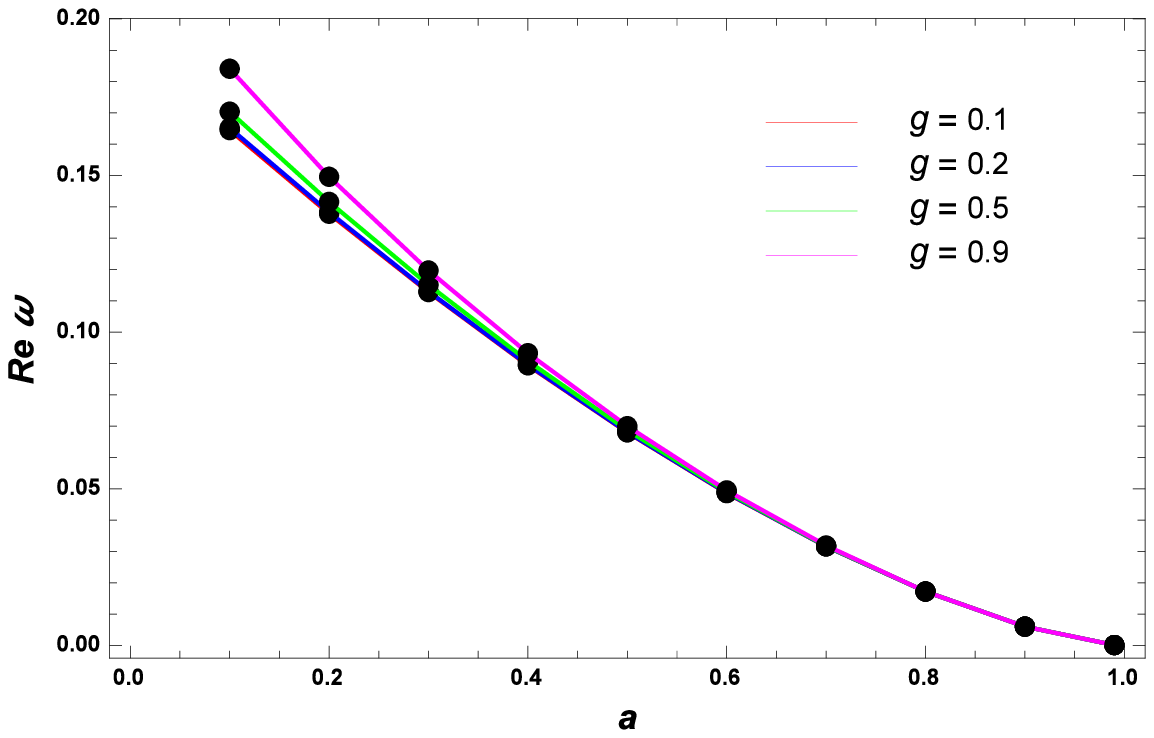}
\includegraphics [width=.5\linewidth]{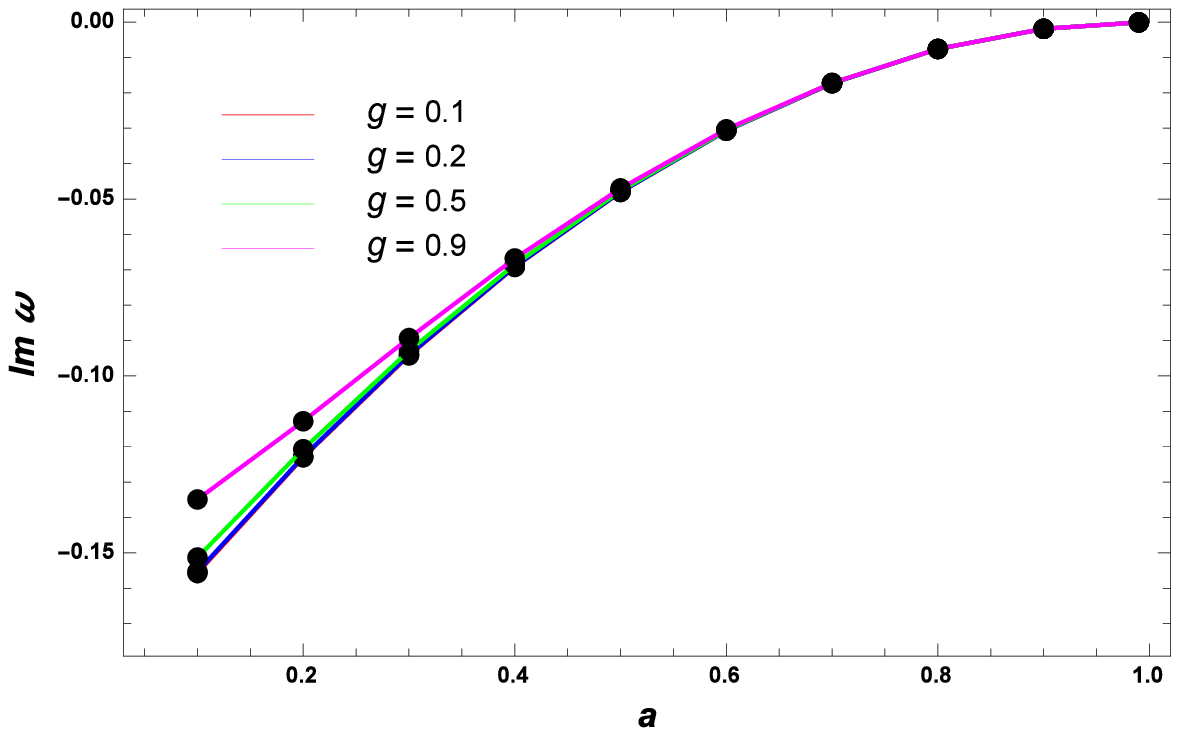}\\
\includegraphics [width=.5\linewidth]{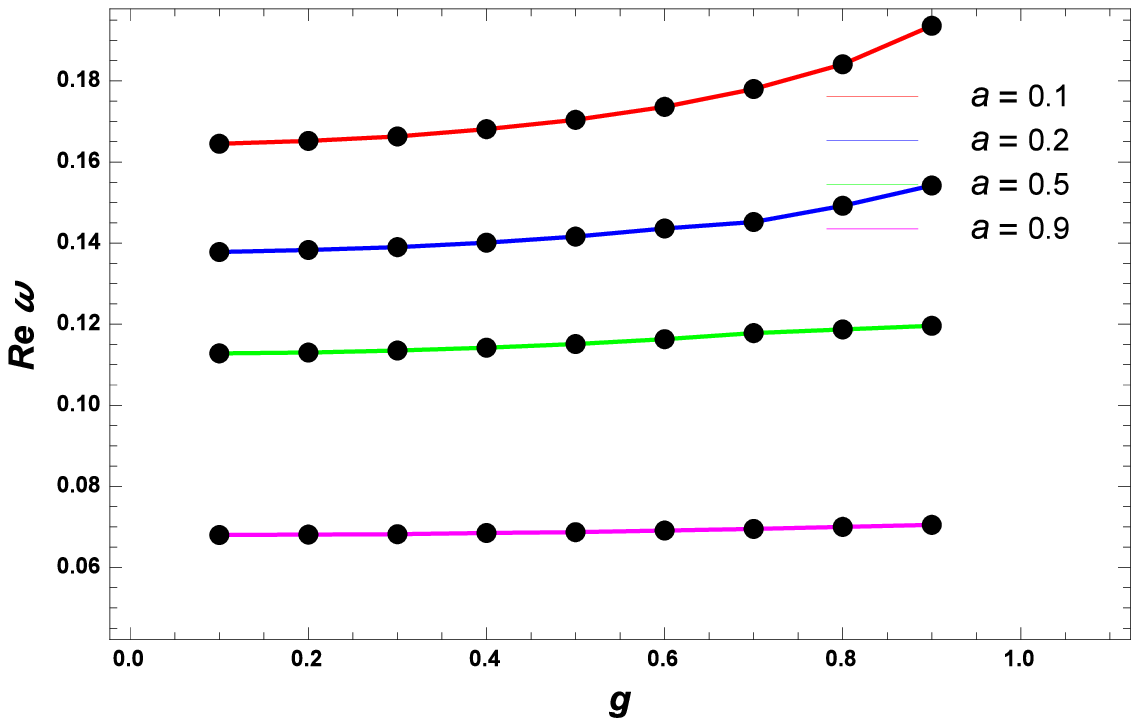}
\includegraphics [width=.5\linewidth]{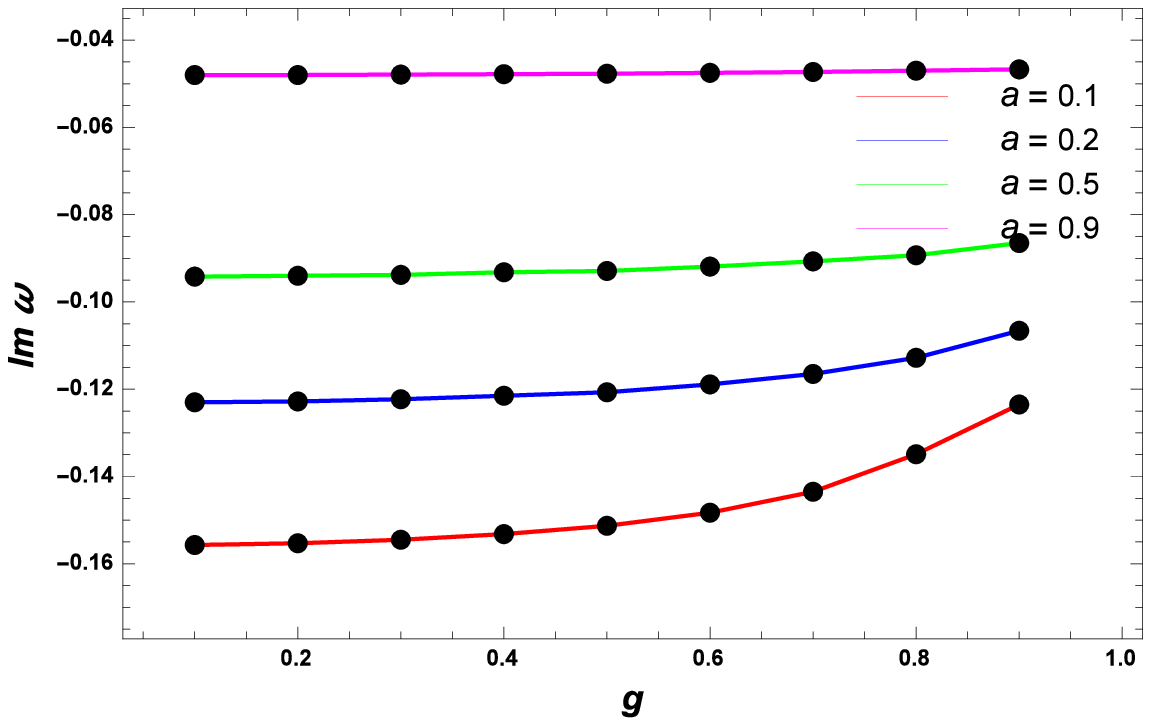}
\end{tabular}
\caption{The real (left panel)  and imaginary part (right panel) of QNMs for different values of  magnetic monopole charge  $(g)$  for a fixed  value of $M$. }
\label{sh}
\end{figure*}

The numerical values of the real and imaginary parts of QNFs are presented in Tab. \ref{tr14} and \ref{tr15} and plotted in Fig. 3 for different values of the BH parameters.  The  negative values of the imaginary part of the QNMs  confirm  that the modes of the obtained BH are stable.

The effect of parameters ($a, g$) on the behavior of QNMs and QNFs are depicted in Fig. 2. In this Fig. 2, we notice that the real part of the QNMs decreases with $a$ while the imaginary part   first increases (becomes less negative) very sharply and then increases slowly (almost constant) with $a$.

\begin{table}[ht]
 \begin{center}
 \begin{tabular}{| l | l   | l   | l   |  l|   }
\hline
            \hline
  \multicolumn{1}{|c|}{ } &\multicolumn{1}{c|}{$g=0.1$}  &\multicolumn{1}{c|}{$g=0.2$}  &\multicolumn{1}{c|}{$g=0.3$} &\multicolumn{1}{|c|}{$g=0.4$}\\
  \hline
    \multicolumn{1}{|c|}{ $a$} &\multicolumn{1}{c|}{$\omega=\omega_R+i \omega_I$}  &\multicolumn{1}{c|}{$\omega=\omega_R+i \omega_I$}  &\multicolumn{1}{c|}{$\omega=\omega_R+i \omega_I$} &\multicolumn{1}{|c|}{$\omega=\omega_R+i \omega_I$}\\
            \hline
            \,\,\,\,\,0.1 ~~  &~~0.1645 - 0.1557  $i$~~  & ~~0.1652 - 0.1553 $i$~~ & ~~0.1663 - 0.1545 $i$~~ & ~~0.1681 - 0.1532 $i$~~ \\            
            \,\,\,\,\,0.2~~ &~~0.1378 - 0.1230 $i$~~ & ~~0.1383 - 0.1228 $i$~~ & ~~0.1390 - 0.1223 $i$~~ &  ~~0.1401 - 0.1215 $i$~~      \\
            \,\,\,\,\,0.3~~ &~~0.1128 - 0.0942 $i$~~  & ~~0.1130 - 0.0940 $i$~~ & ~~0.1135 - 0.09381 $i$~~ &  ~~0.1142 - 0.0933 $i$~~    \\
            \,\,\,\,\,0.4~~ &~~0.0894 - 0.0692 $i$~~ & ~~0.0896 - 0.0691 $i$~~ & ~~0.0899 - 0.0690 $i$~~ &    ~~0.0903 - 0.0688 $i$~  \\
            \,\,\,\,\,0.5~~ &~~0.0680 - 0.0480 $i$~~ & ~~0.0681 - 0.0480 $i$~~ & ~~0.0682 - 0.0479 $i$~~ &    ~~0.0685 - 0.0478 $i$~  \\
            \,\,\,\,\,0.6~~ &~~0.0486 - 0.0307 $i$~~ & ~~0.0487 - 0.0307 $i$~~ & ~~0.0488 - 0.0307 $i$~~ &    ~~0.0488 - 0.0307 $i$~  \\
            \,\,\,\,\,0.7~~ &~~0.0316 - 0.0173 $i$~~ & ~~0.0316 - 0.0173 $i$~~ & ~~0.0316 - 0.0173 $i$~~ &    ~~0.0316 - 0.0172 $i$~  \\
            \,\,\,\,\,0.8~~ &~~0.0172 - 0.0076 $i$~~ & ~~0.0172 - 0.0076 $i$~~ & ~~0.0172 - 0.0076 $i$~~ &    ~~0.0172 - 0.0076 $i$~  \\
            \,\,\,\,\,0.9~~ &~~0.0060 - 0.0019 $i$~~ & ~~0.0060 - 0.0019 $i$~~ & ~~0.0060 - 0.0019 $i$~~ &    ~~0.0060 - 0.0019 $i$~  \\
             \,\,\,\,\,0.99~~ &~~0.0001 - 0.00001 $i$~~ & ~~0.0001 - 0.00001 $i$~~ & ~~0.0001 - 0.00001 $i$~~ &    ~~0.0001 - 0.00001 $i$~  \\
            \hline 
\hline
        \end{tabular}
            \caption{The numerical values of QNMs for various values of  CoS parameter, $a$ and magnetic monopole charge, $g$ for a fixed value of $M=1$ (where $l=1$ and $n=0$).}
\label{tr14}
    \end{center}
\end{table}

\begin{table}[ht]
 \begin{center}
 \begin{tabular}{| l | l   | l   | l   |  l|   }
\hline
            \hline
  \multicolumn{1}{|c|}{ } &\multicolumn{1}{c|}{$g=0.5$}  &\multicolumn{1}{c|}{$g=0.6$}  &\multicolumn{1}{c|}{$g=0.7$} &\multicolumn{1}{|c|}{$g=0.8$}\\
  \hline
    \multicolumn{1}{|c|}{ $a$} &\multicolumn{1}{c|}{$\omega=\omega_R+i \omega_I$}  &\multicolumn{1}{c|}{$\omega=\omega_R+i \omega_I$}  &\multicolumn{1}{c|}{$\omega=\omega_R+i \omega_I$} &\multicolumn{1}{|c|}{$\omega=\omega_R+i \omega_I$}\\
            \hline
            \,\,\,\,\,0.1 ~~  &~~0.1704 - 0.1513  $i$~~  & ~~0.1736 - 0.1483 $i$~~ & ~~0.1780 - 0.1435 $i$~~ & ~~0.1841 - 0.1349 $i$~~ \\            
            \,\,\,\,\,0.2~~ &~~0.1416 - 0.1207 $i$~~ & ~~0.1436 - 0.1189 $i$~~ & ~~0.1462 - 0.1165 $i$~~ &  ~~0.1496 - 0.1128 $i$~~      \\
            \,\,\,\,\,0.3~~ &~~0.1151 - 0.0929 $i$~~  & ~~0.1163 - 0.0919 $i$~~ & ~~0.1178 - 0.0907 $i$~~ &  ~~0.1197 - 0.0893 $i$~~    \\
            \,\,\,\,\,0.4~~ &~~0.0908 - 0.0685 $i$~~ & ~~0.0915 - 0.0681 $i$~~ & ~~0.0923 - 0.0675 $i$~~ &    ~~0.0933 - 0.0668 $i$~  \\
            \,\,\,\,\,0.5~~ &~~0.0687 - 0.0477 $i$~~ & ~~0.0691 - 0.0475 $i$~~ & ~~0.0695 - 0.0473 $i$~~ &    ~~0.0700 - 0.0470 $i$~  \\
            \,\,\,\,\,0.6~~ &~~0.0490 - 0.0306 $i$~~ & ~~0.0491 - 0.0305 $i$~~ & ~~0.0493 - 0.0304 $i$~~ &    ~~0.0495 - 0.0303 $i$~  \\
            \,\,\,\,\,0.7~~ &~~0.0317 - 0.0172 $i$~~ & ~~0.0317 - 0.0172 $i$~~ & ~~0.0318 - 0.0172 $i$~~ &    ~~0.0319- 0.0172 $i$~  \\
            \,\,\,\,\,0.8~~ &~~0.0172 - 0.0076 $i$~~ & ~~0.0172 - 0.0076 $i$~~ & ~~0.0172 - 0.0076 $i$~~ &    ~~0.0172 - 0.0076 $i$~  \\
            \,\,\,\,\,0.9~~ &~~0.0060 - 0.0019 $i$~~ & ~~0.0060 - 0.0019$i$~~ & ~~0.0060 - 0.0019 $i$~~ &    ~~0.0060 - 0.0019 $i$~  \\
             \,\,\,\,\,0.99~~ &~~0.0001 - 0.00001 $i$~~ & ~~0.0001 - 0.00001 $i$~~ & ~~0.0001 - 0.00001 $i$~~ &    ~~0.0001 - 0.00001$i$~  \\
            \hline 
\hline
        \end{tabular}
            \caption{The numerical values of QNMs for various values of  CoS parameter, $a$ and magnetic monopole charge, $g$ for a fixed value of $M=1$ (where $l=1$ and $n=0$).}
\label{tr15}
    \end{center}
\end{table}

\section{Results and Conclusions}

In this article, we considered an exact BH solution of Rodrigue et.el \cite{Rodrigues:2022zph,Rodrigues:2022rfj}, when gravity is minimally coupled to NED  and CoS source. The size of the black hole (outer) horizon decreases with  an increase in magnetic charge, $g$, and increases with CoS parameter, $a$. We studied the photon sphere radii and QNM's in the eikonal limits. The results showed that the photon sphere radius and shadow radius increases with the CoS parameter and opposite behavior with  a magnetic charge. It means that the effect of CoS parameters and magnetic charge are opposite to each other. We also see that the size of the shadow image is bigger than its horizon radius and photon sphere radius.


We employed the well-known connection between the
shadow radii \cite {Singh:2022dth,k1,k2,k3,car1} and QNMs. The real part of the QNMs increases  with the CoS parameter and decreases with the magnetic charge, which means that the QNM oscillates faster. The imaginary part of the QNM is negative, which means that our BH solution is stable. The imaginary part of the  QNM increases with both the CoS parameter and magnetic charge and approaches zero at a very large value of the CoS parameter, which means that the QNMs decayed slower. It would be interesting to study the optical features and mimicker behavior for  other solutions such as charged accelerating AdS BHs,  $f(R)$ gravity coupled with NED black bounce solutions, regular EGB black holes, $f(Q)$ gravity, and wormhole solutions in CoS background.

\begin{acknowledgements}  
  The work of BKV is supported by UGC fellowship. DVS thanks to the DST-SERB project (grant no. EEQ/2022/00824).

\end{acknowledgements}


\end{document}